\documentclass[10pt,letterpaper,twocolumn]{article}
\usepackage[english]{babel}
\usepackage{graphicx}

\newcommand{\alt}{\mathbin{\lower 3pt\hbox
   {$\rlap{\raise 5pt\hbox{$\char'074$}}\mathchar"7218$}}}
\newcommand{\agt}{\mathbin{\lower 3pt\hbox
   {$\rlap{\raise 5pt\hbox{$\char'076$}}\mathchar"7218$}}}

\begin{document}

\setcounter{footnote}{0}
\setcounter{equation}{0}
\setcounter{figure}{0}
\setcounter{table}{0}

\title{\large\bf General form of DMPK equation}

\author{\small I. M. Suslov  \\
\small P.L.Kapitza Institute for Physical Problems,  \\
\small 119334 Moscow, Russia  \\
\small E-mail: suslov@kapitza.ras.ru\\
 {}\\
\parbox{120mm}{\footnotesize \,The
Dorokhov--Mello--Pereyra--Kumar (DMPK) equation,
using in the analysis of quasi-one-dimensional systems
and describing
evolution of diagonal elements of the many-channel
transfer matrix,
is derived under minimal assumptions on the properties
of channels.  The general equation is of the diffusion type
with a tensor character of the diffusion coefficient and finite
values of non-diagonal components. We suggest three different
forms of the diagonal approximation, one of which reproduces
the usual DMPK equation and its generalization suggested by
Muttalib and co-workers. Two other variants lead to equations of
the same structure, but with different definitions of entering
them parameters. They contain additional terms, which are
absent in the first variant.  } }

\date{}
\maketitle

\textwidth 6.4 in
\textheight 8.5 in

\setcounter{footnote}{0}
\setcounter{equation}{0}
\setcounter{figure}{0}
\setcounter{table}{0}

\begin{center}
{\bf 1. Introduction}
\end{center}

The Dorokhov--Mello--Pereyra--Kumar (DMPK) equation
\cite{1,2,3,4} is an efficient instrument for investigation of
quasi-1D disordered systems  (see the paper \cite{5}
for a review). It describes  evolution of diagonal elements of
the many-channel  transfer matrix due to increasing the
system length. The DMPK equation is obtained from the maximum
entropy principle  (assuming the maximal
randomness consistent with the symmetry restrictions) and
conceptually close to the random matrix theory by Wigner and
Dyson \cite{6}. It is determined by one parameter (the system
size in units of the correlation length) and manifests
universality specific for the metallic state. The
DMPK equation is equivalent to the super-symmetric sigma-model
\cite{7}, derived from the microscopic Hamiltonians
\cite{8,9,10}, but  allows to work
with distributions of physical quantities.  Solution of the DMPK
equation \cite{3,4} reproduces the universal fluctuations of
conductance and quantum corrections to it obtained from
diagrammatic calculations \cite{11,12}.

In principle, the transfer matrix approach underlining
the DMPK equation is not restricted by the quasi-1D geometry.
Considering a system of $N$ coupled  one-dimensional
chains and arranging the chains in accordance with
symmetry of a $d$-dimensional lattice, one can construct the
systems of higher dimensionality. However, assumptions
underlining the DMPK
equation lead to statistical equivalence
of chains and eliminate all information on the topology
of space in the transverse directions. As a result, the DMPK
equation cannot be used for study of the Anderson transition
and
is restricted by the metallic phase in the corresponding
$d$-dimensional space. In the localized regime of a
$d$-dimensional system, the DMPK equation is not adequate even
for the quasi-1D geometry: it predicts the minimal Lyapunov
exponent to be of order  $1/N$, while the reasonable
microscopic models lead to the result $O(1)$ \cite{13,14}. The
latter can be understood easily in the regime of strong
localization, when conductance is determined by one resonant
trajectory and the system in fact becomes strictly
one-dimensional\,\footnote{\,The simple algorithm for
construction of resonant trajectories is given in Footnote 4
of the paper \cite{15}. In the strongly localized regime, the
contributions of resonant trajectories to conductance  are
scattered exponentially, and the latter is dominated
by the most transparent channel. }.

It should be clear that in the general case
assumptions used in the DMPK equation should be relaxed; in
particular, it is necessary for describing universality
arising near the Anderson transition.  Derivation of the most
general form of the DMPK equation is a problem, realized by
scientific community \cite{3,4,14,16} and admitted to be of
fundamental significance.  One of the
possible generalizations was suggested by Muttalib and co-workers
\cite{16}--\cite{17}.

We show below that the equation of the DMPK type may be
derived under minimal assumptions on the properties of channels.
Generally, this equation is  of the diffusion type
with a tensor character of the diffusion coefficient and finite
values of non-diagonal components. We consider three different
forms of the diagonal approximation, one of which reproduces
the usual DMPK equation and its generalization suggested
in \cite{16}--\cite{17}. Two other variants lead to equations of
the same structure, but with different definition of their
parameters. They contain the additional term, which is
absent in the first variant and
turned out to be
very actual in the
recent research of the conductance distribution  \cite{15}.

\begin{center}
{\bf 2. Main concepts}
\end{center}

\begin{figure}
\centerline{\includegraphics[width=3.3 in]{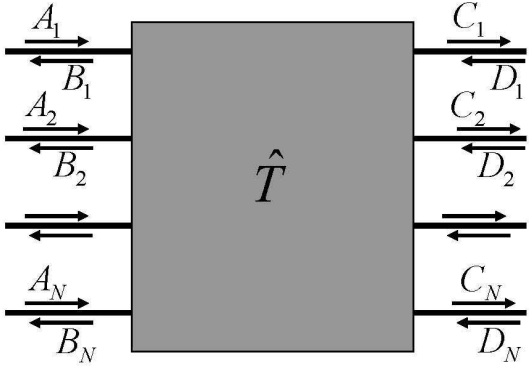}}
\caption{
The many-channel transfer matrix $\hat T$ relates the
amplitudes of the plane waves on the left  ($A_n$, $B_n$)
and on the right ($C_n$, $D_n$) of a scatterer.  }
\label{fig1}
\end{figure}

Considering the system as a set of $N$ coupled
one-dimensional chains, we can thought of it as a "black box"
 with attached ideal contacts in the form of
isolated $1D$ conductors\,\footnote{\,It means that a
concept of "channels" is used in the real space representation,
which removes all problems related with evanescent modes
\cite{17a}.}. Then the system can be treated as an effective
scatterer and described by the transfer matrix  $\hat T$,
relating the amplitudes of waves on the left
($A_n e^{ikx}+B_n e^{-ikx}$ in  $n$-th channel) and on the right
($C_n e^{ikx}+D_n e^{-ikx}$) of the scatterer  (Fig.1):
$$
\left ( \begin{array}{cc} A \\ B \end{array} \right)\, =
\hat T \left ( \begin{array}{cc} C \\ D \end{array} \right) \,
= \left ( \begin{array}{cc} T_{11} & T_{12} \\ T_{21} & T_{22}
\end{array} \right)\,
\left ( \begin{array}{cc} C \\ D \end{array} \right) \,,
\eqno(1)
$$
where $A$, $B$, $C$, $D$ are vectors with components
$A_n$, $B_n$, $C_n$, $D_n$. In the vector notation, the form
of Eq.\,1 does not depend on the number of channels, while the
transfer matrix is divided naturally  into four blocks;
it allows a parametrization \cite{3,18}
$$
 \hat T= \left ( \begin{array}{cc} u_1 & 0 \\
         0 & v_1 \end{array} \right)\,
\left ( \begin{array}{cc} \sqrt{1+\lambda} & \sqrt{\lambda} \\
\sqrt{\lambda} & \sqrt{1+\lambda} \end{array} \right)\,
	 \left ( \begin{array}{cc} u & 0 \\
         0 & v \end{array} \right)\,,
\eqno(2)
$$
where $u$, $v$, $u_1$, $v_1$ are unitary matrices, and
$\lambda$ is a diagonal matrix with the positive elements
$\lambda_i$, which are eigenvalues of the Hermitian matrix
$T_{12}T^+_{12}$. In the presence of time reversal invariance
the additional relations arise \cite{2,3}
$$
v=u^*\,,\qquad v_1=u_1^*\,,
\eqno(3)
$$
which as a rule are of no significance for the following.

Of the main interest are parameters  $\lambda_i$, which in
particular determine the conductance
$$
g=\sum\limits_{i} \frac{1}{1+\lambda_i}\,
\eqno(4)
$$
(in the Economou--Soukoulis definition  \cite{19,37}). The DMPK
equation describes evolution of their mutual distribution
function  $P\left(\lambda_1, \lambda_2, \ldots, \lambda_N
\right) \equiv P\{\lambda\}$ with increasing  the length
$L$ of the system
$$
\frac{\partial P\{\lambda\}}{\partial L} = \alpha\,
\sum\limits_{i} \frac{\partial }{\partial \lambda_i}
\left[\,\lambda_i (1+\lambda_i)\, J\{\lambda\}
\frac{\partial }{\partial \lambda_i}\,
\frac{P\{\lambda\}}{J\{\lambda\}} \,\right]
\eqno(5)
$$
$$
J\{\lambda\} = \prod\limits_{i<j} |\lambda_i-\lambda_j|^\beta
\,,
$$
where $\beta=1$ for the orthogonal ensemble (usual systems
with a random potential), $\beta=2$ for the unitary ensemble
(systems in the strong magnetic field), $\beta=4$ for the
symplectic ensemble (system with the strong spin-orbit
interaction); parameter $\alpha$ has a sense of the inverse
correlation length of the quasi-1D system. The quantity
$J\{\lambda\}$ is well-known from the random matrix theory
$\cite{6}$ and arises from the Jacobian of transformation
$$
\prod\limits_{ij} d H_{ij} = J\{\lambda\}
\,\tilde J\{Q\}\,\prod\limits_{i} d\lambda_i \prod\limits_{ij}
dQ_{ij}
\eqno(6)
$$
when integration over elements of the matrix $\hat H$ is
replaced by integration over its eigenvalues $\lambda_i$
and the elements of diagonalizing matrix  $\hat Q$
($\hat H=\hat Q^{-1}\hat \Lambda\hat Q$). In fact,
$J\{\lambda\}$ is the distribution function of levels, if
they are contained in the restricted interval with the
periodic boundary conditions (the Dyson circular
ensemble). In actual applications, the distribution
$P\{\lambda\}$ includes the additional factor, providing
localization of the spectrum in a finite interval and practically
not affecting the distribution of close levels. Analogously,
the distribution  $P\{\lambda\}=J\{\lambda\}$ is a formal
solution of  equation (5) but does not satisfy the
normalization condition; an additional factor  becomes
inevitable, whose evolution is described by the DMPK
equation. Exact solution of Eq.5 for $\beta=2$ shows \cite{19a}
that the additional factor is not reduced to a smooth envelope
but essentially changes the distribution $P\{\lambda\}$, making it
different from  $J\{\lambda\}$ even on the local level;
correlations of $\lambda_i$ are determined by the
Jacobian $J\{\lambda\}$ only at the initial stage of evolution,
when all  $\lambda_i$ are small.   In the context of
generalizations of the DMPK equation this circumstance acquires a
deep sense  (see Footnote 9 in Sec.\,5).

In a strictly one-dimensional system we have  $J\{\lambda\}=1$
and equation (5) reduces to the form
$$
\frac{\partial P(\lambda)}{\partial L} =
\alpha\,\frac{\partial}{\partial \lambda}
\left[\,\lambda(1+\lambda)\,\frac{\partial P(\lambda)}{\partial \lambda}
\,\right] \,, \eqno(7)
$$
and $\lambda$ coincides with the Landauer resistance  $\rho$
\cite{20}; such equation was derived in many papers
\cite{21}--\cite{25}. Recently it was shown by the present
author  \cite{15} that the general evolution equation in
1D systems has a form
$$
\frac{\partial P(\lambda)}{\partial L} =
\alpha\,\frac{\partial}{\partial \lambda}
\left[-\gamma(1+2\lambda) P(\lambda) +
\lambda(1+\lambda)\frac{\partial P(\lambda)}{\partial \lambda}
\,\right]  ,
\eqno(8)
$$
and the additional term, specified by  parameter  $\gamma$,
is physically significant: its incorporation in the Shapiro
scheme  \cite{26} allows to explain all essential
features in the conductance distribution, which was
impossible on the basis of  (7). This term disappears in the
random phase approximation and is naturally not reproduced
by equation  (5), based on the analogous assumptions.
However, this term is also not predicted by the generalized
DMPK equation suggested by Muttalib and co-workers
\cite{16}--\cite{17}
$$
\frac{\partial P\{\lambda\}}{\partial L} = \alpha
\sum\limits_{i} K_{ii} \frac{\partial }{\partial \lambda_i}
\left[\,\lambda_i (1+\lambda_i)\, J_i\{\lambda\}
\frac{\partial }{\partial \lambda_i}\,
\frac{P\{\lambda\}}{J_i\{\lambda\}} \,\right]
\eqno(9)
$$
$$
J_i\{\lambda\} = \prod\limits_{j<k}
|\lambda_j-\lambda_k|^{\beta^i_{jk}}\,,\qquad
\beta^i_{jk}=2K_{jk}/K_{ii}
$$
and containing as parameters the elements  $K_{ij}$ of a
certain matrix $\hat K$.\,\footnote{\,The $i$ dependence
of $J_i\{\lambda\}$ was unnoticed in \cite{16a} (see Appendix
$B$), but practically  it is not very actual \cite{40}.}
This fact clearly demonstrates that attempts
of generalization of the DMPK equation are not sufficiently
advanced and do not reproduce all essential contributions. This
point was the main motivation of the present paper.

\begin{center}
{\bf 3. Idea of derivation}
\end{center}

Derivation of the evolution equation is based on the
relation
$$
\hat T_{L+\Delta L}=\hat T_{L}\,\hat T_{\Delta L} \,,
\eqno(10)
$$
where $\hat T_{\Delta L}$ is a matrix close to the unit one.
The form of the DMPK equation depends on statistical
properties of the parameters $\epsilon_k$, determining
deviations of $\hat T_{\Delta L}$ from the unit matrix. These
parameters can be divided into two groups: for the first of them
$$
\langle\epsilon_k \rangle \ne 0\,,
\eqno(11)
$$
while for the second
$$
\langle\epsilon_k \rangle= 0\,,\qquad
\langle\epsilon^2_k \rangle\ne  0\,.
\eqno(12)
$$
The presence of parameters (12) is necessary for arising
of an equation of the diffusion type: since there is no effect
in the first order in  $\epsilon_k$, all calculations should
be made in the second order, and this is a reason for
appearing of the second derivatives which are characteristic for
the diffusion equation. The general strategy consists in
averaging only over parameters (12), and making no assumptions
relative to parameters (11).

To obtain the most general form of the DMPK equation, we
should distinguish the category of quantities, for which the
property (12) is not a model assumption but is an inherent
property,
following from their nature. Such quantities are well
known and related with a diagonal disorder. Consider the
Schroedinger equation with a random potential, which is
defined on the lattice sites by a set of independent
random quantities  $V_n$ (as in
the Anderson model). Variables $V_n$ should have
identical distributions to provide a spatial homogeneity in
average.  If the mean value $\langle V_n \rangle$ is finite,
then it is equal for all  $n$ and can be excluded by
a shift of the origin of energy $E$,
since a random
potential enters in the combination  $V_n-E$.  Thus, we can
accept without a loss of generality
$$
\langle V_n \rangle= 0\,,\qquad
\langle V^2_n \rangle =W^2\,,
\eqno(13)
$$
as it is made in almost all
theoretical papers.

\begin{figure}
\centerline{\includegraphics[width=3.2 in]{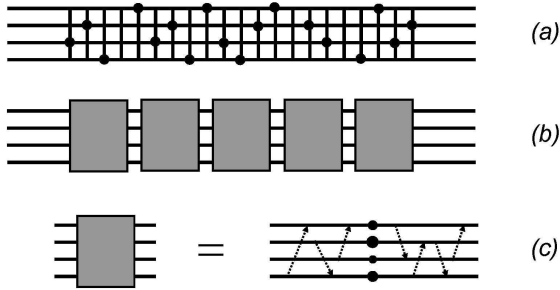}}
\caption{\small
(a) A typical quasi-1D system is a bar, cutout of the
$d$-dimensional lattice, with randomly located impurities
inserted in it.  (b) The system can be divided into a sequence
of effective scatterers, whose transfer matrices are multiplied.
(c) Each scatterer provides a partial reflection of the
incident waves and mixing of channels; these two processes
can be imagined as somewhat separated in space.  } \label{fig2}
\end{figure}

This
point can be used in the following manner. A typical
quasi-1D system is a bar cutout of a $d$-dimensional lattice
and containing randomly located impurities (Fig.\,2,a).
We can divide it into a series of effective scatterers
containing a lot of lattice sites (Fig.\,2,b). Each
scatterer should provide existence of two effects,
(a) a partial reflection of the incident waves, and (b) mixing
of channels. It is convenient to imagine these two processes as
slightly separated
in space (Fig.\,2,c), so that there is a
region where waves are reflected without mixing of channels,
and there are two regions where channels are mixed for the
transmitted and back-scattered waves but no reflection
occurs. Such assumption is not very essential, since we suppose
nothing on the degree of separation
and it can be purely symbolic.
In fact, the construction in Fig.\,2,c corresponds to the
canonical representation (2) of the transfer matrix:
it is easy to see that the middle matrix in (2) provides
a reflection of waves without mixing of channels, while the
right and the left matrices provide mixing of channels
without reflection of waves.

The middle part of the effective scatterer  (Fig.\,2,c)
can be described by the transfer matrix
$$
\left ( \begin{array}{cc} 1-i\epsilon & -i\epsilon \\
i\epsilon & 1+i\epsilon \end{array} \right)\,,
\eqno(14)
$$
corresponding to the diagonal disorder created
by point scatterers on the independent one-dimensional chains,
so $\epsilon$ is a diagonal matrix with real elements
$\epsilon_k$, possessing the property (12). Extracting from (14)
the factors, not related with scattering, we can accept the
following representation for matrix $\hat T_{\Delta L}$
$$
\hat T_{\Delta L}= \left ( \begin{array}{cc}\!\! w_1\! & \!0\!\! \\
         \!\!0 \! & \!w_2\!\! \end{array} \right)\,
\left ( \begin{array}{cc} \!\!\sqrt{1+\epsilon^2} & -i\epsilon \\
i\epsilon & \!\sqrt{1+\epsilon^2}\!\! \end{array} \right)\,
	 \left ( \begin{array}{cc} \!\!w_3\! &\! 0\!\! \\
       \! \! 0 \! & \! w_4\! \!\end{array} \right),
\eqno(15)
$$
where $w_1$, $w_2$, $w_3$, $w_4$ are matrices close to the unit
one and elements  $\epsilon_k$ are small. Accepting the canonical
representation (2) for  $\hat T_{L}$ and composing the product
(10), one can see that matrices $w_1$, $w_2$ lead to a
small renormalization of matrices  $u$ and $v$, which can
be neglected\,\footnote{\,We do not use the canonical
representation (2) in (15), since in this case matrices $w_i$
do not tend to the unit one in the limit $\epsilon\to 0$,
leading to a finite renormalization of  $u$ and $v$.
The use of (14) as a middle matrix of (15) leads to the more
tremendous calculations.}. It is clear that we should
compose the product
$$
\hat T'=
\left ( \begin{array}{cc} \sqrt{1+\lambda} & \sqrt{\lambda} \\
\sqrt{\lambda} & \sqrt{1+\lambda} \end{array} \right)\,
\left ( \begin{array}{cc} u & 0 \\
         0 & v \end{array} \right)
         \,\cdot \qquad
$$
$$\qquad\qquad\qquad\cdot\,
\left ( \begin{array}{cc} \sqrt{1+\epsilon^2} & -i\epsilon \\
i\epsilon & \sqrt{1+\epsilon^2} \end{array} \right)\,
\eqno(16)
$$
and reduce it to the canonical form (2).

The assumption of a diagonal disorder for the middle part
of an effective scatterer  (Fig.\,2,c) is not very essential.
Indeed, the notion of weak scatterers is inevitable in derivation
of the differential equation; in the opposite case only the
finite difference equation is possible. Having in mind a
description of the Anderson transition, we should work near the
band edge of the ideal crystal, since only there a weak
disorder is compatible with localization in higher dimensions.
Then the de Broglie wavelength and the mean free path are large in
comparison with the atomic spacing and the wave function
envelope changes slowly. It allows to introduce the coarse
description, dividing the system into blocks, small in
comparison with the wavelength but containing a lot of the lattice
sites, and considering these blocks as the new lattice sites.  In
the result of such procedure practically any short-ranged random
potential reduces to the diagonal Gaussian disorder. Universality
arising near the Anderson transition as in other critical
phenomena \cite{28,29} leads to equivalence of its description
near the band edge and in the band center.

\begin{center}
{\bf 4. General evolution equation}
\end{center}

Let describe the general scheme of deriving the evolution
equation, while the calculation details can be found in
Appendix $A$. Parameters $\lambda'_i$ of the matrix $\hat
T'$ can be found as eigenvalues of the Hermitian
"Hamiltonian"
$H=T_{12}T^+_{12}$, where
$$
T_{12} = \sqrt{1+\lambda}\,u\,(-i\epsilon)+
\sqrt{\lambda}\, v \,\sqrt{1+\epsilon^2}
\,, \eqno(17)
$$
and can be presented as functions of $\lambda_i$
($\lambda'_i=f_{i}\{\lambda\}$) in the form of expansion over
$\epsilon$. Composing the distribution function of $\lambda'_i$,
we have
$$
P_{L+\Delta L}\{\lambda'\}= \int \prod\limits_i d\lambda_i
P_{L}\{\lambda\} \prod\limits_i
\delta\left(\lambda'_i-f_{i}\{\lambda\} \right) \cdot
$$
$$
\cdot
P(\epsilon)
\,P(u,v) \,d\epsilon\, du\, dv \,.
\, \eqno(18)
$$
Making a change of variables $y_i=f_{i}\{\lambda\}$, one
can replace integration over $\lambda_i$  by integration over
$y_i$
$$
\prod\limits_i d\lambda_i = I\{y\} \prod\limits_i dy_i
\,, \eqno(19)
$$
while the inverse relations  $\lambda_i=g_{i}\{y\}$ are
found by iterations in $\epsilon$. Integration over $y_i$
removes the $\delta$-functions and leads to the result
$$
P_{L+\Delta L}\{\lambda\}= \int  I\{\lambda\}
P_{L}\{g_i\{\lambda\}\} P(\epsilon) P(u,v)\, d\epsilon\, du\,
dv \,. \eqno(20)
$$
In calculation of the Jacobian $I\{y\}$ one discovers that its
diagonal elements are of order unity, while non-diagonal elements
are of order  $\epsilon^2$, so in fact it reduces to the
product of diagonal elements. Substituting for $I\{\lambda\}$ and
$g_i\{\lambda\}$ their expansions in $\epsilon$
and expanding (20) to the second order, we produce averaging
according to  $\langle\epsilon_k \rangle= 0$,
$\langle\epsilon_k \epsilon_{k'} \rangle
=\langle\epsilon^2 \rangle\, \delta_{kk'}$ and set\,\footnote{\,In
coarsening of description discussed in the end of Sec.\,3,
the variances of individual scatterers are added and their sum
$\langle\epsilon^2 \rangle$ is proportional to the volume;
it gives a linear dependence on $\Delta L$ in the quasi-1D
geometry. With such definition, the parameter  $\alpha$ appears
to be of the order of the inverse mean free path.}
$\langle\epsilon^2 \rangle\equiv \alpha \Delta L$. As a result
$$\!\!
\frac{\partial P\{\lambda\}}{\partial L}\! =\alpha\!
\sum\limits_{i} \frac{\partial }{\partial \lambda_i}\!
\left[G_i\{\lambda\}P\{\lambda\}\! + \!
\sum\limits_j
F_{i\! j}\{\lambda\}
\frac{\partial P\{\lambda\} }{\partial \lambda_j}\,\!
\right]
\eqno(21)
$$
where the following functions of $\lambda_i$ are introduced
(the primes near the summation signs indicate the absence of
terms with $j=i$)
$$
F_{ij}\{\lambda\} = \frac{1}{2}
\sqrt{\lambda_i (1+\lambda_i) \lambda_j (1+\lambda_j)}
 \,A_{ij} \,,
$$
$$
G_{i}\{\lambda\} = (1+2\lambda_i)\left(\frac{1}{2}A_{ii}-1\right)
+ \qquad
\eqno(22)
$$
$$+
\sqrt{\lambda_i (1+\lambda_i)}\, {\sum\limits_j}'
\frac{1+2\lambda_j}{4\sqrt{\lambda_j(1+\lambda_j)}}\, A_{ij} -
\tilde G_{i}\{\lambda\}  \,,
$$
$$
\tilde G_{i}\{\lambda\} = {\sum\limits_j}'
\frac{\lambda_i(1+\lambda_j)\, B_{ij} +\lambda_j(1+\lambda_i)
\,C_{ij}} {\lambda_i-\lambda_j} \,+
$$
$$\qquad\qquad\qquad
   + {\sum\limits_j}'
\frac{\sqrt{\lambda_i (1+\lambda_i) \lambda_j(1+\lambda_j)}
        } {\lambda_i-\lambda_j} \, D_{ij}
$$
with a definition of matrices
$$
B_{ij}=\sum\limits_k \overline{|v_{ik}|^2 |u_{jk}|^2}\,,
\qquad
C_{ij}=\sum\limits_k \overline{|u_{ik}|^2 |v_{jk}|^2}\,,
$$
$$
D_{ij}=-\sum\limits_k \left(\overline{v_{ik} v_{jk}
u^*_{ik}u^*_{jk}} +\overline{v^*_{ik} v^*_{jk}
u_{ik}u_{jk}}\right)\,,
$$
$$
A_{ij}=\sum\limits_k
\left(\overline{u_{ik} u^*_{jk}v^*_{ik} v_{jk}}
+\overline{u^*_{ik} u_{jk} v_{ik} v^*_{jk}} -
\right.
\eqno(23)
$$
$$  \qquad\quad   \left.
-\overline{u_{ik} u_{jk} v^*_{ik} v^*_{jk}}
-\overline{u^*_{ik} u^*_{jk} v_{ik} v_{jk}} \right)\,.
$$
Equation (21)  is
the most general form of the
DMPK equation: in its derivation we did not use
any assumptions on the statistical properties of matrices  $u$
and $v$, and  they even are not obliged to be random. The right
hand side of (21) is a sum of full derivatives, which provides
the conservation of the total probability.

\begin{center}
{\bf 5. Diagonal forms}
\end{center}

Equation (21) is of the diffusion type, with a tensor character
of the diffusion coefficient and finite non-diagonal
components. In the general form it is rather complicated and
hardly suitable for a constructive analysis; so consider its
possible simplifications.

Equation (21) is simplified radically, if we assume the
diagonal form for matrices  $A_{ij}$ and $D_{ij}$
$$
A_{ij}=A_i\delta_{ij}\,, \qquad D_{ij}=D_i \delta_{ij}\,.
\eqno(24)
$$
We accept also $B_{ij}=C_{ij}\equiv K_{ij}$, since the
statistical properties of matrices  $u$ and $v$ are usually
identical.  Then equation (21) reduces to the form  (see
Appendix $B$)
$$
\frac{\partial P\{\lambda\}}{\partial L} = \alpha\,
\sum\limits_{i}
\frac{1}{2} A_{i} \frac{\partial }{\partial \lambda_i}
\left[\,-\gamma_i (1+2\lambda_i)\, P\{\lambda\} +
\phantom{\frac{P\{\lambda\}}{J_i\{\lambda\}}}
\right.
$$
$$\qquad
\left.+
\lambda_i (1+\lambda_i)\, J_i\{\lambda\}
\frac{\partial }{\partial \lambda_i}\,
\frac{P\{\lambda\}}{J_i\{\lambda\}} \,\right]
\eqno(25)
$$
$$
\gamma_i=(2K_{ii}-A_i)/A_i \,,\qquad
$$
$$
J_i\{\lambda\} = \prod\limits_{j<k}
|\lambda_j-\lambda_k|^{\beta^i_{jk}}\,,\qquad
\beta^i_{jk}=4K_{jk}/A_{i}\,,
$$
which reproduces Eq.\,8 in the one-channel case.  Conditions
for realization of the diagonal approximation can be easily
analyzed for the unitary ensemble, when matrices  $u$ and $v$ are
averaged independently. If a unitary matrix $u$ is restricted
by real values of its elements, then it turns into the orthogonal
matrix $\tilde u$; to restore the unitary matrix we should add
to the elements of  $\tilde u$ the appropriate phase factors.
Producing the same manipulations with the matrix $v$, we set
$$
u_{lk}=\tilde u_{lk} e^{i\varphi_{lk}}\,,\qquad
v_{lk}=\tilde v_{lk} e^{i\phi_{lk}}\,
\eqno(26)
$$
and obtain after substitution to (23)
$$\!
B_{ij}=\sum\limits_k \left\langle|\tilde v_{ik}|^2
|\tilde u_{jk}|^2 \right\rangle \,, \quad
C_{ij}=\sum\limits_k \left\langle|\tilde u_{ik}|^2
|\tilde v_{jk}|^2 \right\rangle,
$$
$$\!\!
D_{ij}=\!-2\sum\limits_k \left\langle\tilde v_{ik}
\tilde v_{jk} \tilde u_{ik} \tilde u_{jk} \cos(\phi_{ik}\! +\!
\phi_{jk}\!- \!\varphi_{ik}\!-\!\varphi_{jk} \right)\rangle
\eqno(27)
$$
$$\!\!
A_{ij}=4\sum\limits_k \left\langle\tilde v_{ik} \tilde v_{jk}
\tilde u_{ik} \tilde u_{jk}
 \sin(\varphi_{ik}\!-\!\phi_{ik}) \sin(\varphi_{jk}\! -\! \phi_{jk}
\right)\rangle
$$
If matrices $\tilde v$ and $\tilde u$ are completely
random, while  phases $\varphi_{ik}$ and $\phi_{ik}$
have nonuniform distributions, then products
$\tilde v_{ik} \tilde v_{jk}$, $\tilde u_{ik} \tilde u_{jk}$
are averaged to zero for $i\ne j$,
%
%
providing the diagonal
approximation (24) where $A_i$ and $K_{ij}$ are
independent, and the trivial result is valid for $K_{ij}$
(see Eq.\,28 below). Contrary, if matrices  $\tilde v$ and
$\tilde u$ are not sufficiently random, but phases  $\varphi_{ik}$
and $\phi_{ik}$ are completely stochastic, then we have another
diagonal approximation with nontrivial values of  $K_{ij}$ and
relation $A_i=2K_{ii}$; as a result, the terms with
$\gamma_i$ turn to zero and Eq.\,25 reduces to the variant (9),
suggested by Muttalib et al \cite{16}--\cite{17}. Finally,
if both  $\tilde v$, $\tilde u$, and  $\varphi_{ik}$,
$\phi_{ik}$ are completely random, then  averaging occurs
over the unitary group (see Appendix $B$ in \cite{5})  and
leads to the results
$$
K_{ij}=\sum\limits_k \left\langle|v_{ik}|^2 \right\rangle
\left\langle|u_{jk}|^2 \right\rangle =\frac{1}{N}
\qquad\mbox{\rm and} \quad \beta^i_{jk} =2 \,,
 \eqno(28)
$$
$$
K_{ij}=\sum\limits_k \left\langle| u_{ik}|^2
| u_{jk}|^2 \right\rangle =\frac{1+\delta_{ij}}{N+1}\,
\quad\mbox{\rm and} \quad \beta^i_{jk} =1
\eqno(29)
$$
for the unitary and the orthogonal ensembles correspondingly,
so equation  (9) transforms to the usual DMPK equation
(5).\,\footnote{\,For the orthogonal ensemble, the first
diagonal approximation (25) is not realized.}


Let us discuss the third variant of the diagonal
approximation, which we consider as the most actual. It was
argued in \cite{15,30}, that for the correct definition of
conductance of a finite system it is useful to introduce
semi-transparent boundaries, separating the system from
the ideal leads attached to it. In the limit of weak
transparency one obtains universal equations, independent on the
way how the contact resistance of the reservoir is excluded
\cite{31} (all formulas of the Landauer type
\cite{32}--\cite{36} reduce in this limit to the variant by
Economou--Soukoulis \cite{19,37}), which then can be
extrapolated to transparency of order unity. Such definition
is surely referred to the system under consideration
(and not to the composed system "sample+ideal leads") and
provides the infinite value of conductance for an ideal system
 \cite{30}.

Suppose that weakly-transparent boundaries are created by
point scatterers inserted in one-dimensional chains
attached to the system (Fig.\,2,a); then its
transfer matrix $\hat T$  transforms to
$\hat T_0\hat T \hat T_0$, i.e.
$$
\left ( \begin{array}{cc} \!\! 1\!-\!i\kappa & -i\kappa \\
         i\kappa & 1\!+\!i\kappa \!\!\end{array} \right)
\left ( \begin{array}{cc} T_{11} & T_{12}\! \\
T_{21} & T_{22}\! \end{array} \right)
	 \left ( \begin{array}{cc} \!\! 1\!-\!i\kappa & -i\kappa \\
         i\kappa & 1\!+\!i\kappa \!\!\end{array} \right),
\eqno(30)
$$
where $\kappa$ is a diagonal matrix.
Reducing
(30) to the canonical form (2), one has in the main
approximation for large $\kappa$
$$
u_1 \sqrt{1+\lambda}\, u = - \kappa \tilde T \kappa\,,
$$
$$
u_1 \sqrt{\lambda}\, v = - \kappa \tilde T \kappa\,,
$$
$$
v_1 \sqrt{\lambda}\, u =  \kappa \tilde T \kappa\,,
$$
$$
v_1 \sqrt{1+\lambda}\, v =  \kappa \tilde T \kappa\,,
\eqno(31)
$$
where $\tilde T=T_{11}-T_{12}+T_{21}-T_{22}$.
Since the unitary matrices $u$, $v$, $u_1$, $v_1$ have
restricted elements, then $\lambda\sim \kappa^4$ and
$1+\lambda$ can be replaced by $\lambda$; then (31) gives
$$
u=v, \quad u_1=-v_1\,\quad \mbox{\rm for}\,\quad  \kappa
\to \infty\,.
\eqno(32)
$$
For large $\lambda_i$ equations (21--23) reduce to the
form analogous to (25), but with another definition of
$K_{ij}$ (see Appendix $B$), $K_{ij}=(B_{ij}+C_{ij}+D_{ij})/2$.
Substitution of (32) into (23) gives $K_{ij}\to 0$,
$A_{ij}\to 0$ in the $\kappa\to\infty$ limit. For large, but
finite $\kappa$ the small deviations of $v$ from $u$ should
be taken into account, setting
$$
v=u e^{ih}\,,
\eqno(33)
$$
where $h$ is the Hermitian matrix with small elements.
Substituting (33) into (23) and expanding to the second
order in $h$, one has
$$
2 K_{ij}=\sum\limits_k \left\langle |u_{ik}|^2 |u_{jk}|^2
\left(|h_{ik}|^2 + |h_{jk}|^2+
\right.\right.
\qquad\qquad\qquad
$$
$$\qquad\qquad\qquad\qquad
+\left.\left.
h_{ik}h_{jk}+ h^*_{ik}h^*_{jk}\right) \right\rangle \,,
$$
$$
A_{ij}=\sum\limits_k \left\langle |u_{ik}|^2 |u_{jk}|^2
\left(h_{ik}h_{jk} +h^*_{ik}h^*_{jk}+
\right.\right.\qquad\qquad
$$
$$
\qquad\qquad\qquad\qquad+\left.\left.
h^*_{ik}h_{jk}+h_{ik}h^*_{jk}\right) \right\rangle\,.
\eqno(34)
$$
It is  easy to see, that $A_{ii}=2K_{ii}$ independently of
the $h_{ik}$ statistics  (in fact, it follows from the general
expressions (23)).  For large  $\kappa$, the quantities $h_{ik}$
are small in magnitude, but there are no other restrictions on their
statistics. It is natural to think that $h_{ik}$
fluctuate randomly and their fluctuations are independent of
$u_{ik}$.\,\footnote{\,If matrix $u$ contains a dependence
on  $h$, then this dependence manifests only in terms of order
$h^3$, which are neglected in (34). } Then pair products
$h_{ik}h_{jk}$, $h^*_{ik} h_{jk}$, $\ldots$ with $i\ne j$
are averaged to zero, and the matrix  $A_{ij}$
becomes diagonal. As a result, equation (21) accepts the
form
$$
\frac{\partial P\{\lambda\}}{\partial L} = \alpha\,
\sum\limits_{i}
 K_{ii} \frac{\partial }{\partial \lambda_i}
\left[\,-\gamma_i (1+2\lambda_i)\, P\{\lambda\} +
\phantom{\frac{P\{\lambda\}}{J_i\{\lambda\}}}
\right.
$$
$$
\left.+
\lambda_i (1+\lambda_i)\, J_i\{\lambda\}
\frac{\partial }{\partial \lambda_i}\,
\frac{P\{\lambda\}}{J_i\{\lambda\}} \,\right]
\eqno(35)
$$
$$
J_i\{\lambda\} = \prod\limits_{j<k}
|\lambda_j-\lambda_k|^{\beta^i_{jk}}\,,\qquad
\beta^i_{jk}=2K_{jk}/K_{ii}
$$
$$
\gamma_i=(1-\sum_j K_{ij})/K_{ii} \,,\qquad
K_{ij}=(B_{ij}+C_{ij}+D_{ij})/2 \,
$$
and has the same structure as (25),  but with different
definition of parameters. Since  $K_{ij}$ are small for large
$\kappa$, parameters $\gamma_i$ are surely finite and
large in magnitude.

The first two diagonal approximations look somewhat
artificial. If matrices  $u$ and $v$ are completely
random, then we return to the initial equation (5). If
$u$ and $v$ are not sufficiently random, then a tendency to the
non-diagonal situation arises: we do not see  serious grounds
why $\tilde u_{ij}$ should be more random than $\varphi_{ij}$ or
vice versa.  Contrary,  the third variant of a diagonal
approximation looks quite natural: existence of weakly
transparent boundaries restricts mutual fluctuations of $u$
and $v$, but beyond these restrictions they are considered as
completely random. Simultaneously, all situation with the definition
of conductance becomes logically consistent.

It is well-known  \cite{2,5}, that equation (5) is easily
solved in the limit of large $L$, when parameters $\lambda_i$
are large and obey hierarchy
$\lambda_1\gg\lambda_2\gg\ldots\gg\lambda_N$; then $J\{\lambda\}$
reduces to the product of powers of  $\lambda_i$ and equation
(5) splits into  $N$ independent equations. Applying the same
procedure to equation  (35), we find the independent Gaussian
distributions for quantities  $x_i=\ln\lambda_i$ defined by
their first two moments:
$$
\langle x_i\rangle = \alpha L \left[(2\gamma_i+1) K_{ii} +
2\sum\limits^N_{j=i+1} K_{ij}\right] \,, \qquad
$$
$$
\sigma^2_i=\langle x_i^2\rangle -\langle x_i\rangle^2 = 2\alpha L
\,K_{ii}\,, \eqno(36)
$$
which for $\gamma_i=0$ coincides with results of  \cite{14,16a}.
In the approximation of equivalent channels one can set
$\alpha K_{ii}=\tilde \alpha$, $\beta_{ij}=\beta$,
$\gamma_i=\gamma$, and equation (35) is determined by
three parameters $\tilde\alpha L$, $\beta$, $\gamma$; in
particular,
$$
\frac{2\langle x_i\rangle}{\sigma^2_i}=
2\gamma+1 + \beta(N-i)
\eqno(37)
$$
and parameters $\beta$, $\gamma$ can be easily estimated
from numerical data on Lyapunov exponents (see e.g.
\cite{38a,38}). One can see from formula (32) of the paper
\cite{38} that relation  $\sigma^2_i=2\langle x_i\rangle$
for the minimal exponent ($i=N$ in our notation) is
valid in the metallic regime but violated in other cases;
hence the parameter  $\gamma$ is finite beyond the
metallic phase.\,\footnote{\,The formula (4.5) of the paper
\cite{17} contains the more general expression for
$\langle x_i\rangle$, reflecting a violation of the
strong hierarchy of $\lambda_i$ in the quasi-3D geometry;
it reduces to results of \cite{14,16a} in the
$L\to\infty$ limit for fixed $N$, which is a proper limit for a
definition of the Lyapunov exponents. Probably, in conditions
of the paper  \cite{17} the matrix  $D_{ij}$ was diagonal with
nonzero elements $D_{ii}$; as a result, finiteness of $\gamma_i$
was compensated by redefinition of $K_{ii}$ and did not affect
the quality of fitting on the basis of formula  (4.5).}


As clear from derivation, the structure of equation  (35)
is the same for the unitary and orthogonal ensembles;
correspondingly, $\beta$ becomes a free parameter not
related with the Wigner--Dyson values, and in the general case
transforms to
a matrix\,\footnote{\,At first glance, for
non-integer $\beta$ we meets with violation of the repulsion
law for two nearest levels at their anomalous approaching.
In fact  (see discussion after formula  (6)),
correlation of levels if determined by the Jacobian
$J\{\lambda\}$ only in the region of small  $L$, where
$\beta$  coincides with its Wigner--Dyson value.}. It is
clear that the "pure" Wigner--Dyson ensembles loose their
actuality beyond the metallic phase, and in particular
are not adequate for description of the Anderson
transition. The latter circumstance  is not accounted for
in the existent versions of the sigma-models \cite{8,9,10},
which are equivalent to the simplest equation  (5) and require
modification for  incorporation of
the discussed generalizations.
The only exclusion is the situation for  $d=2+\epsilon$, where
universality arising near the critical point approximately
corresponds to universality specific for the metallic phase,
which is adequately described by equation  (5).  It provides
validity of results in the main
$\epsilon$-approximation but remains the open question
on their validity in  higher orders.

\begin{center}
{\bf 6. Conclusion}
\end{center}

In the present paper we derive the DMPK equation under minimal
assumptions on the properties of channels. It is of the diffusion
type with a tensor character of the diffusion coefficient and
nonzero off-diagonal components. We suggest three variants of the
diagonal approximation, one of which reproduces the
usual DMPK equation and its generalization suggested in
 \cite{16}--\cite{17}. Two other variants lead to equations of
the same structure and contain additional terms specified by
parameters  $\gamma_i$.

The most general form of the DMPK equation, given by Eq.\,21,
probably is not very actual: it should be used as a starting
point for
new statistical hypotheses, which were
adequate for  description of the Anderson transition. The
methods used in numerical experiments allow to calculate matrices
$u$ and $v$ \cite{17a}, and analyzing their statistical
properties establish the form of matrices  $A_{ij}$, $B_{ij}$,
$C_{ij}$, $D_{ij}$.  Numerical analysis  undertaken
in the context of equation (9)
\cite{17,40}\,\footnote{\,It should be noted that the present
paper clarifies the conditions for validity of equation (9); in
particular, self-averaging of  $K_{ij}$, discussed in details
by the authors of
\cite{40}, in fact is of no significance.}, points out the
realization of the diagonal approximation and deviation of
parameters $\beta^i_{jk}$ from their Wigner--Dyson values;
a finiteness of parameters $\gamma_i$ follows from Eq.32 of
\cite{38}. It is desirable to continue such analysis on the basis
of the general expressions (23). On the other hand, mathematical
methods developed for analysis of the usual DMPK equation
\cite{3,4,5}, can be used for deriving more general
relations; existence of large parameters $\gamma_i$ may provide
new possibilities.

\begin{center}
{\it Appendix A.} Derivation of the evolution equation
 \end{center}

Parameters $\lambda'_i$  of the matrix $\hat T'$ can be found
as eigenvalues of the Hermitian "Hamiltonian"
$H=T_{12}T^+_{12}$
(see (17)), which has the matrix elements\,\footnote{\,All
calculations are produced to the second order in $\epsilon$.
The imaginary unit $i$ enters in the several expressions as
a factor and is easily distinguished from indices.}
$$
H_{ij}=\lambda_i \delta_{ij}+V_{ij}  \,,
$$
$$
V_{lj} = i \sum\limits_k \epsilon_k
\left[\sqrt{\lambda_l(1+\lambda_j)} \,v_{lk} u^*_{jk}
-\right. \qquad\qquad\qquad
$$
$$\left.\qquad\qquad\qquad\qquad
     -\sqrt{(1+\lambda_l)\lambda_j} \,u_{lk} v^*_{jk} \right] +
\eqno(A.1)
$$
$$
        + \sum\limits_k \epsilon^2_k
\left[\sqrt{(1+\lambda_l)(1+\lambda_j)}\, u_{lk} u^*_{jk}
     +\sqrt{\lambda_l \lambda_j}\, v_{lk} v^*_{jk} \right]
\,.
$$
Eigenvalues  $\lambda'_i$ of the matrix  $H$ are calculated by
the usual perturbation theory
$$
\lambda'_i=\lambda_i+V_{ii}+
{\sum\limits_j}' \frac{V_{ij}V^*_{ij}}{\lambda_i-\lambda_j}
\,, \eqno(A.2)
$$
and have a form of expansion in  $\epsilon_k$
$$
\lambda'_i= f_{i}\{\lambda\} =
\lambda_i+ \sqrt{\lambda_i (1+\lambda_i)}
\sum\limits_k A^i_{k}\epsilon_k +
$$
$$+
\sum\limits_{kk'} C^i_{kk'}\{\lambda\} \epsilon_k\epsilon_{k'}
\,, \eqno(A.3)
$$
with the coefficients
$$
A^l_{k}=i\left(v_{lk} u^*_{lk}-u_{lk} v^*_{lk}\right)\,,
$$
$$
B^i_{k}\{\lambda\}= (1+\lambda_i) |u_{ik}|^2 +\lambda_i
|v_{ik}|^2     \,,
$$
$$
C^i_{kk'}\{\lambda\} = B^i_{k}\{\lambda\} \delta_{kk'}+
\phantom{\frac{P\{\lambda\}}{J_i\{\lambda\}}}
$$
$$+
{\sum\limits_j}'
\frac{\lambda_i(1+\lambda_j)B_{ijkk'} +
(1+\lambda_i)\lambda_j C_{ijkk'}}{\lambda_i-\lambda_j} \,  +
$$
$$
+{\sum\limits_j}'
\frac{\sqrt{\lambda_i (1+\lambda_i) \lambda_j(1+\lambda_j)}}
{\lambda_i-\lambda_j} \, D_{ijkk'}  \,,
\, \eqno(A.4)
$$
$$
B_{ijkk'}=v_{ik} v^*_{ik'} u^*_{jk} u_{jk'} \,,
$$
$$
C_{ijkk'}=u_{ik} u^*_{ik'} v^*_{jk} v_{jk'} \,,
$$
$$
D_{ijkk'}=-v_{ik} u^*_{ik'} u^*_{jk} v_{jk'}
          -u_{ik} v^*_{ik'} v^*_{jk} u_{jk'}  \,.
$$
Composing the distribution (18) and making a change of variables
$y_i=f_{i}\{\lambda\}$, one comes to Eq.\,20, where the inverse
functions $\lambda_i=g_{i}\{y\}$ are found by iterations in
$\epsilon_k$
$$
\lambda_i=g_{i}\{y\} =y_i-
\sqrt{y_i(1+y_i)}\sum\limits_k A^i_{k}\epsilon_k +
$$
$$+
{\textstyle\frac{1}{2}}(1+2y_i)\sum\limits_{kk'} A^i_{k}A^i_{k'}
\epsilon_k \epsilon_{k'}
-\sum\limits_{kk'} C^i_{kk'}\{y\} \epsilon_k\epsilon_{k'}
\,. \eqno(A.5)
$$
Integration over  $y_i$ removes the $\delta$-functions and
leads to the result (20). The Jacobian matrix $I\{y\}$ has
diagonal elements of order unity and off-diagonal elements
of order $\epsilon^2$,
$$
\frac{\partial\lambda_i}{\partial y_i}=
1-\frac{(1+2y_i)}{2\sqrt{y_i(1+y_i)}}
\sum\limits_k A^i_{k}\epsilon_k +
$$
$$+
\sum\limits_{kk'} A^i_{k}A^i_{k'}
\epsilon_k \epsilon_{k'}  -
\sum\limits_{kk'} \frac{\partial C^i_{kk'}\{y\}}{\partial y_i}
 \epsilon_k\epsilon_{k'}    \,,
$$
$$
\frac{\partial\lambda_i}{\partial y_j}=
-\sum\limits_{kk'} \frac{\partial C^i_{kk'}\{y\}}{\partial y_j}
 \epsilon_k\epsilon_{k'}\,\qquad (j\ne i)
\,, \eqno(A.6)
$$
so its determinant reduces to the product of diagonal
elements. It is calculated according to the scheme
$$
\prod\limits_i (1+a_i \epsilon + b_i \epsilon^2)\approx
1+\sum\limits_i a_i \epsilon + \sum\limits_i b_i \epsilon^2
+{\textstyle\frac{1}{2}}
{\sum\limits_{ij}}' a_i a_j \epsilon^2
\,, \eqno(A.7)
$$
and takes a form
$$
I\{\lambda\}= 1+\sum\limits_k R_{k}\{\lambda\}\epsilon_k +
\sum\limits_{kk'} S_{kk'}\{\lambda\} \epsilon_k\epsilon_{k'}
\,, \eqno(A.8)
$$
where
$$
R_k\{\lambda\}=-\sum\limits_i\frac{(1+2\lambda_i)}
{2\sqrt{\lambda_i(1+\lambda_i)}}\, A^i_{k}  \,,
\eqno(A.9)
$$
$$
S_{kk'}\{\lambda\}=\sum\limits_i \left( A^i_{k} A^i_{k'}-
\frac{\partial C^i_{kk'}\{\lambda\}}{\partial \lambda_i} \right)
+
$$
$$
+\frac{1}{8} {\sum\limits_{ij}}'
\frac{(1+2\lambda_i) (1+2\lambda_j)}
{\sqrt{\lambda_i(1+\lambda_i)\lambda_j(1+\lambda_j)}}\,
 A^i_{k} A^i_{k'} \,.
$$
Now we can make the expansion
$$
P_L\{ g_{i}\{\lambda\}\} =P_L\{\lambda_i+\Delta \lambda_i\}=
P_L\{\lambda\} +
$$
$$+
\sum\limits_i \frac{\partial P_L\{\lambda\}}{\partial \lambda_i}
       \Delta \lambda_i
+\frac{1}{2}\sum\limits_{ij} \frac{\partial^2
P_L\{\lambda\}}{\partial \lambda_i \partial \lambda_j}
\Delta \lambda_i \Delta \lambda_j
\,, \eqno(A.10)
$$
where
$$
\Delta\lambda_i= - \sqrt{\lambda_i (1+\lambda_i)}
\sum\limits_k A^i_{k}\epsilon_k +
\sum\limits_{kk'} L^i_{kk'}\{\lambda\} \epsilon_k\epsilon_{k'}
\,,
$$
$$
L^i_{kk'}\{\lambda\} = {\textstyle\frac{1}{2}}
(1+2\lambda_i) A^i_{k} A^i_{k'}
-C^i_{kk'}\{\lambda\} \,.
\eqno(A.11)
$$
Substituting ($A.8-A.11$) into (20) and averaging according
$\langle\epsilon_k \rangle= 0$,
$\langle\epsilon_k \epsilon_{k'} \rangle =\alpha \Delta
L \delta_{kk'}$, one has
$$
\frac{\partial P\{\lambda\}}{\alpha\partial L} = \,
P\{\lambda\} \sum\limits_k \overline{S_{kk}\{\lambda\}} +
$$
$$
+\sum\limits_{i} \frac{\partial P\{\lambda\}}{\partial \lambda_i}
\sum\limits_k \left( \overline{L^i_{kk}\{\lambda\}}
-\sqrt{ \lambda_i(1+\lambda_i)}\,
         \overline{A^i_{k} R_k\{\lambda\}} \right)
$$
$$
+\frac{1}{2}\sum\limits_{ij} \frac{\partial^2
P_L\{\lambda\}}{\partial \lambda_i \partial \lambda_j}
\sqrt{\lambda_i (1+\lambda_i) \lambda_j (1+\lambda_j)}\,
\sum\limits_k \overline{A^i_{k} A^j_{k}}
  \,,
\eqno(A.12)
$$
which can be transformed to Eqs.\,21--23.

\begin{center}
{\it Appendix B.}  Simplification of equation (21).
\end{center}

In the diagonal approximation (24) equation (21) accepts
a form
$$
\frac{\partial P\{\lambda\}}{\partial L} = \alpha\,
\sum\limits_{i}  \frac{\partial }{\partial
\lambda_i} \left[\,G_i\{\lambda\}P\{\lambda\} +
\phantom{\frac{P\{\lambda\}}{J_i\{\lambda\}}}
\right.
$$
$$
+\left.
\frac{1}{2} A_i\lambda_i (1+\lambda_i)
\frac{\partial P\{\lambda\} }{\partial \lambda_i}\,
 \,\right] \,,
\eqno(B.1)
$$
$$
G_{i}\{\lambda\} = (1+2\lambda_i) \frac{A_i-2}{2}
- {\sum\limits_j}'\,
\frac{2\lambda_i\lambda_j+\lambda_i+\lambda_j}
{\lambda_i-\lambda_j} K_{ij}\,.
$$
The sum over $j$ can be transformed using the
identity \cite{2}
$$
{\sum\limits_j}'\frac{K_{ij}}{\lambda_i-\lambda_j}=
\frac{\partial \ln{J\{\lambda\}}}{\partial \lambda_i} \,,
\quad J\{\lambda\} = \prod\limits_{i<j}
|\lambda_i-\lambda_j|^{K_{ij}}\,,
\eqno(B.2)
$$
which is valid for a symmetrical matrix $K_{ij}$. It allows to
simplify the combination
$$
\frac{1}{2}A_i \frac{\partial {P\{\lambda\}}}{\partial \lambda_i}
-2\,\frac{\partial \ln{J\{\lambda\}}}{\partial
\lambda_i}P\{\lambda\}=
$$
$$
=
\frac{1}{2}A_i J_i\{\lambda\}\frac{\partial }{\partial \lambda_i}
\frac{P\{\lambda\}}{J_i\{\lambda\}}
\,,\quad
J_i\{\lambda\}\equiv J\{\lambda\}^{4/A_i}
\eqno(B.3)
$$
and reduce $(B.1)$ to the form (25). If a symmetry requirement
for $K_{ij}$ is ignored in $(B.2)$, then it is easy to
arrive at a false conclusion that  $J_i\{\lambda\}$ are
independent of $i$ and determined by parameters
$\beta_{ij}=4K_{ij}/A_i$.

In the case of weakly transparent boundaries, parameters
 $\lambda_i$ are large and expansions over  $1/\lambda_i$
 are possible with retaining the first two terms;
 in particular,
 $$
\sqrt{\lambda_i (1+\lambda_i) \lambda_j (1+\lambda_j)} \approx
(2\lambda_i\lambda_j+\lambda_i+\lambda_j)/2\,,\qquad
$$
$$
\sqrt{\lambda_i (1+\lambda_i)} \approx
(1+2\lambda_i)/2\,
\eqno(B.4)
$$
and one has in Eq.\,22
$$
\tilde G_{i}\{\lambda\}=
-(1+2\lambda_i) {\sum\limits_j}'K_{ij} +
{\sum\limits_j}'(B_{ij}-C_{ij})/2+
$$
$$
+2\lambda_i(1+\lambda_i) {\sum\limits_j}'
\frac{K_{ij}}{\lambda_i-\lambda_j}\,,
\eqno(B.5)
$$
where $K_{ij}=(B_{ij}+C_{ij}+D_{ij})/2$. Having in mind that
relation $B_{ij}=C_{ij}$ holds usually,
we neglect the second term  in the right hand side, but retain
the symmetric definition for $K_{ij}$. Using $(B.2)$, we can
reduce (21), (22) to a form (35).

\end{document}